\documentclass[manuscript]{aastex}

\shorttitle{A Multi-Epoch SMA Study of the HH 211 Protostellar Jet: Jet Motion and Knot Formation}
\shortauthors{Jhan et al.}

\def\arcsa#1#2{$#1^{\prime\prime}_{^\textrm{.}}#2$}
\def\cirg#1#2{$#1^{\circ}_{^\textrm{.}}#2$}

\begin{document}

\title{A Multi-Epoch SMA Study of the HH 211 Protostellar Jet: Jet Motion and Knot Formation
}

\author{
Kai-Syun Jhan\altaffilmark{1,2} \& Chin-Fei Lee\altaffilmark{1}
}

\affil{Academia Sinica Institute of Astronomy and Astrophysics, P.O. Box 23-141, Taipei 106, Taiwan}

\affil{Graduate Institute of Astronomy and Astrophysics, National Taiwan University, No. 1,
Sec. 4, Roosevelt Road, Taipei 10617, Taiwan}
\altaffiltext{}{E-mail: cflee@asiaa.sinica.edu.tw}

\begin{abstract}
HH 211 is a highly collimated jet with a chain of well-defined knots, powered by a nearby young Class 0 protostar. We have used 4 epochs (2004, 2008, 2010, and 2013) of Submillimeter Array (SMA) archive data to study the properties of the HH 211 jet in SiO (J=8-7). The jet shows similar reflection-symmetric wiggle structures in all epochs. The wiggle structures can all be fitted by an orbiting jet source model that includes a position shift due to proper motion of the jet, indicating that the wiggle propagates along the jet axis. Thus, this suggests the wiggle is indeed due to an orbital motion of the jet source. Proper motions of the knots are measured by using the peak positions of the knots in four epochs, and they are roughly the same and independent of the distance from the central source. The mean proper motion of the knots is $\sim$ \arcsa{0}{087} per year, resulting in a transverse velocity of $\sim$ 114 km s$^{-1}$, about 30\% lower than that measured before. Knots BK2 and BK3 have a well-defined linear velocity structure, with the fast jet material upstream to the slow jet material. The gradient of the velocity structure decreases from knot BK2 to BK3. 
In addition, for each knot, the gradient decreases with time, as the knot propagates away from the central source. These results are both expected if the two knots trace internal shocks produced by a small periodical variation in ejection velocity of the jet.

\end{abstract}


\section{Introduction}

Protostellar jets play an important role in star formation. They are believed to be launched from the innermost parts of accretion disks around
protostars, carrying away excess angular momentum from the disks, allowing the disk material to fall onto the protostars \cite[see the review by][]{2014prpl.conf..451F}.  
The launching mechanism of the jets is still uncertain, but can be constrained with the physical properties (e.g., morphology and kinematics) of the jets.  In particular, since most of the jets have knotty structures (knots), studying the formation of these knots may help us to constrain how the jets are launched.

HH 211 is a well-studied jet consisting of a chain of well-defined knots, and thus it is a good candidate for studying the physical properties of the jet and the formation of
the knots.  It is nearby, located in the IC 348 complex of
Perseus at a distance of $\sim\,$280 pc
\citep{2006ApJ...638..293E,2006AJ....131.1574L}.  The jet is powered by a
low-mass and low-luminosity ($\,\sim\,$3.6 L$_{\bigodot}$, corrected for the
new distance) Class 0 protostar \citep{2005ApJS..156..169F}.  It has been
detected in H$_{2}$ \citep{1994ApJ...436L.189M}, as well as
SiO and CO \citep{1999A&A...343..571G,2006ApJ...636L.141H,2006ApJ...636L.137P,2007ApJ...670.1188L}. It is highly collimated with a small-scale wiggle. 
The wiggle appears to be reflection-symmetric and semi-periodical, and thus is likely due to an orbital motion of the
jet source around a close binary companion \citep{2010ApJ...713..731L}. 
Similar wiggle structure was also observed later in other HH jets \cite[e.g., HH 111 in the infrared in][]{2011ApJ...732L..16N}.
Such an orbiting jet source model has also been proposed to produce the reflection-symmetric wiggle \citep{1998A&A...334..750F,2009ApJ...707L...6R}.
However, it is also possible that the wiggle structure is due to an instability in the jet. 
Therefore, in this paper, we compare the jet structures in SiO (J=8-7) in four epochs to ensure that the wiggle structure is indeed due to an orbital motion of the jet source, instead of an instability.  
 In addition, using more
epochs of observations, we can refine the proper
motion measurement of the knots, and thus the velocity and mass-loss rate of the jet
derived earlier in \citet{2009ApJ...699.1584L}.

%epochs of observations. We can then refine other properties, e.g., the mass-loss rate and the launching radius of the jet.

The formation mechanism of the knots in the jet is still not well determined. The knots seem to be a train of dense clumps produced by episodic ejections. However, current simulations show that even if the mass-loss rate of the jet is constant,
a small periodical variation in ejection velocity can also produce the knots \citep{1990ApJ...364..601R,1993ApJ...413..210S,1997A&A...318..595S,2004ApJ...606..483L}. In these simulations, the knots are formed as the fast jet material catches up with the slow jet material. 
This possibility is supported by previous observations that show a velocity variation (e.g., radial velocity ramps) in the knots along the jet axis, not only in optical \cite[e.g.,][]{2002ApJ...565L..29R,2007AJ....133.1221B} but also in submillimeter wavelengths \cite[e.g.,][]{2009ApJ...699.1584L,2010ApJ...717...58H,2015ApJ...805..186L}.  Notice that the jet can be well mapped in both wavelengths. In order to further support this possibility, we track some well-defined knots of HH 211 in SiO (J=8-7)  in four epochs and check if the evolution of the velocity variation in the knots is consistent with that expected in the simulations. The goal is to determine if the knots are formed because of episodic ejections of jet material or because of a small periodical variation in ejection velocity.

%determine the formation mechanism of the knots by studying their evolution.

\section{Observations}

Archive data of the Submilimeter Array (SMA) were used for our study.
In the archive, four epochs of SiO (J=8-7) observations of HH 211 were found, one in 2004, one in 2008, one in 2010, and one in 2013.
The observations in 2010 and 2013 epochs were carried out in polarization mode.
The array configurations, resolutions, and sensitivities are listed in Table 1. 
These resolutions and sensitivities are sufficient for our study. 
In all epochs, continuum emission of the central source was detected at $\sim$ 340 GHz, allowing us to center the jet for proper motion measurement. 

The data in the  2004, 2008, and 2013 epochs have been reported in \citet{2007ApJ...670.1188L}, \citet{2009ApJ...699.1584L}, and \citet{2014ApJ...797L...9L}, respectively. Here we only describe the data in the 2010 epoch.
In this epoch, the HH 211 jet was observed on January 2010 in the extended configuration. Like other epochs, the SiO line and the 340 GHz continuum were observed using 345 GHz band receivers. The receivers have two sidebands: a lower sideband with frequency ranging from 333.5 to 337.5 GHz, and an upper sideband with frequency ranging from 345.5 to 349.5 GHz. Each sideband has 48 chunks, and each chunk has 128 spectral channels with a  velocity resolution of $\sim$ 0.7 km s$^{-1}$ per channel.
After combining these two sidebands without the line emission portion, we obtained a continuum with a bandwidth of $\sim$ 7.5 GHz centered at $\sim$ 340 GHz. The projected baseline lengths range from 30 to 186 m. The primary beam has a size of $\sim$ 35$\arcsec$.

We used MIR package to calibrate the visibility data. The quasar 3c454.3, quasars 3c84 and J0325+469, and Titan were used as passband, gain, and flux calibrator respectively. The flux uncertainty is $\sim$ 15 \%. We then used the MIRIAD package to image the calibrated visibility data. The calibrated visibility data were used to produce dirty maps and a dirty beam. We CLEANed the maps to obtain CLEAN component maps. The final images were obtained by restoring the CLEAN component maps with a Gaussian beam fitted to the main lobe of the dirty beam. With natural weighting, the synthesized beam has a size of \arcsa{1}{0} $\times$ \arcsa{0}{68}  at a position angle (P.A.) of $\sim$ 63$^{\circ}$. The rms noise level is $\sim$ 0.1 Jy beam$^{-1}$ for the SiO channel map and $\sim$ 1 mJy beam$^{-1}$ for the continuum map. 

\section{Results}
\subsection{Wiggle Structure}
\begin{figure}
\epsscale{.55}
\plotone{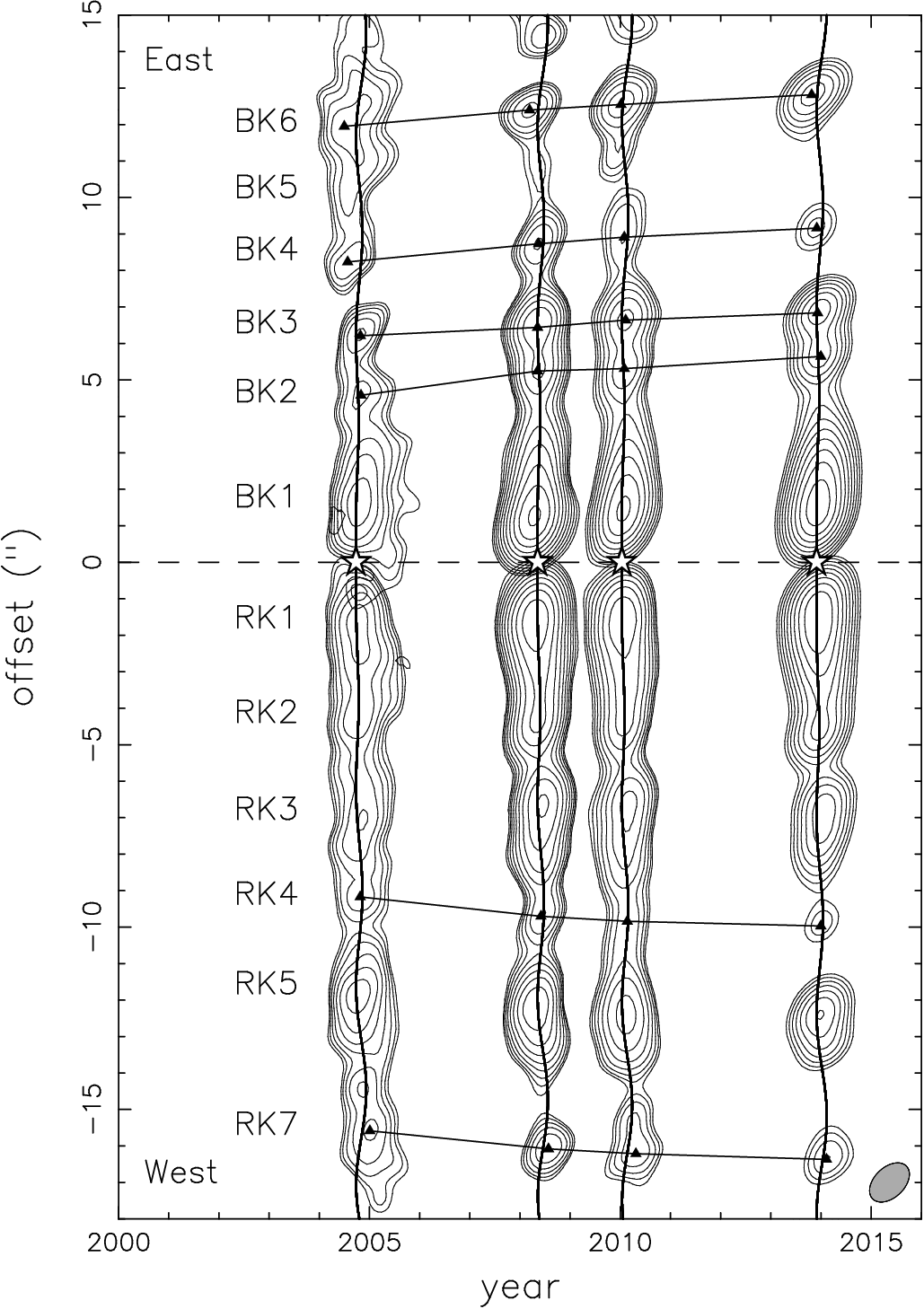}
\caption{SiO maps in different epochs, rotated by \cirg{116}{1} clockwise in order to align the jet with the y-axis. They show the total intensities of SiO emission integrated from -21.2 to 9.2 km s$^{-1}$ (for the blueshifted side) and 9.2 to 47.5 km s$^{-1}$ (for the redshifted side). Contour levels for the 2004 epoch are 0.1$\times$ 1.4$^{n-1}$ Jy beam$^{-1}$, where n = 1,2,3..., while contour levels for other three epochs are 0.05 $\times$ 1.4$^{n-1}$ Jy beam$^{-1}$, where n = 1,2,3.... The asterisks denote the central source position. The triangles mark the peak positions of the knots, and the lines connect the peak position for the same knot in different epochs. The curved lines along the jet structure shows the fits of the wiggle with an orbiting jet source model \citep{2010ApJ...713..731L}. The beam at the bottom-right corner shows the resolution of the maps. \label{siorotate}}
\end{figure}

Figure \ref{siorotate} shows the SiO maps of the HH 211 jet in four epochs. The SiO maps are rotated by \cirg{116}{1} clockwise to align the jet with the y-axis. Knots BK1 to BK6 indicate the blueshifted (eastern) knots, and RK1 to RK5 and RK7 indicate the redshifted (western) knots, as in \citet{2009ApJ...699.1584L}. All the maps are convolved with the same synthesized beam with a size of \arcsa{1}{29} $\times$ \arcsa{0}{83} at a P.A. of $\sim$ 70$^{\circ}$, and then plotted with the same contour levels. Notice that the noise level in the 2004 epoch ( $\sim$ 5 $\times$ 10$^{-2}$ Jy beam$^{-1}$ ) is about twice as those in the other three epochs ( $\sim$ 2 $\times$ 10$^{-2}$ Jy beam$^{-1}$). Therefore, the contour levels of the map in the 2004 epoch start at $\sim$ 0.1 Jy beam$^{-1}$, which corresponds to the third contour level in the other epochs.

The jet shows knotty and wiggle structures on either side of the source (Fig.\,\ref{siorotate}). 
Some knots, BK4, BK5, and RK4, seem to become fainter in the later epochs. 
The wiggle in the 2008 epoch was found to be reflection-symmetric about the central source and semi-periodical \citep{2010ApJ...713..731L}. This kind of wiggle is likely due to an orbital motion of the jet source in a binary system and can be fitted reasonably well by such a model \citep{2010ApJ...713..731L}. This wiggle of the jet is also seen in other epochs, with a position shift due to proper motion of the jet. As shown by the curvy lines in Figure \ref{siorotate}, the wiggles in other epochs can be all fitted well with the orbiting jet source model after including a position shift due to the proper motion of the jet (to be derived later). Therefore, the wiggle propagates along the jet axis and indeed must be due to an orbital motion of the jet source, not due to an instability in the jet.

\subsection{Proper motion}

Since the knots are localized and their structures are similar in different epochs, the peak positions of the knots can be used to estimate the proper motion of the knots and thus the jet. Previously, \citet{2009ApJ...699.1584L} used the 2004 and 2008 epochs to estimate the proper motion. Here, we use two more epochs, 2010 and 2013, providing a total of four epochs spanning over $\sim$ 9 years, to refine the proper motion. In all epochs,  the peaks of the continuum emission of the central source are used to center the jet. Six knots -- BK2, BK3, BK4, BK6, RK4, and RK7-- appear like single knots and thus are used for the proper motion measurement, as in \citet{2009ApJ...699.1584L}. 

In Figure \ref{siorotate}, the triangles mark the peak positions of those knots, and the lines connect the peak positions for the same knots. Fitting the peak positions of the same knots with a linear equation $y=at+b$, [where $y$, $t$, and $a$ stand for the peak position, the epoch of observation, and the slope (i.e., proper motion), respectively], we can obtain the proper motion, $a$, of the knots averaged over the four epochs.

\begin{figure}
\epsscale{1}
\plotone{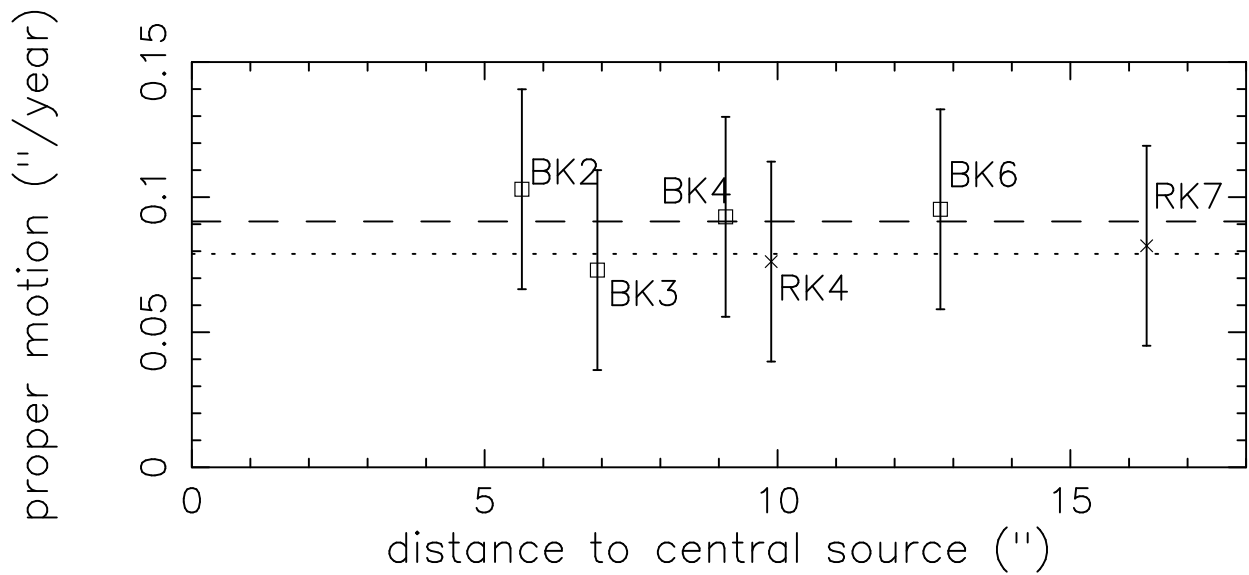}
\caption{The relation of the proper motion and distance of the knots to the central source. The dotted line indicates the mean value of the proper motion on the redshifted side ($\sim$ \arcsa{0}{08} per year), and the dashed line indicates the mean value of the proper motion on the blueshifted side ($\sim$ \arcsa{0}{09} per year). The errorbar has a value of $\pm$ \arcsa{0}{37} per year. \label{fitv}}
\end{figure}

Figure \ref{fitv} shows a plot of the proper motion of the knots  versus the distance of the knots from the central source. 
The proper motion of the knots ranges from $\sim$ \arcsa{0}{07} to $\sim$ \arcsa{0}{1} per year. 
Since the knots are not spatially resolved, the errorbar in the proper motion here is estimated by dividing the possible error in the peak position of the knots by the time span ($\sim$ 9 years) over the four epochs. Assuming that
the possible error in the peak position is less than one third of the beam size along the jet axis, which is $\sim$ \arcsa{0}{3},
the errorbar in the proper motion is $\sim$ \arcsa{0}{037} per year.
The mean proper motion on the redshifted side ($\sim$ \arcsa{0}{08} per year as indicated as the dotted line in Figure \ref{fitv}) is slightly smaller than that on the blueshifted side ($\sim$ \arcsa{0}{09} per year as indicated as the dashed line in Figure \ref{fitv}).  Since the jet is almost in the plane of the sky,
this difference may suggest the jet to have different velocities on each side, as seen in, e.g., Sz 102 \citep{2014ApJ...786...99L}. However, this difference could also be due to an uncertainty in the center position of the jet, which is defined by the peak position of the continuum emission of the central source.

The mean proper motion of all the knots is $\sim$ \arcsa{0}{087} $\pm$ \arcsa{0}{037} per year, resulting in a transverse velocity of $\sim$ 114 $\pm\,50$ km s$^{-1}$. The proper motion here is $\sim$ 30$\%$ lower than that measured in \citet{2009ApJ...699.1584L} using the first two epochs, which was $\sim$ \arcsa{0}{13} $\pm$ \arcsa{0}{04} per year. This difference can be seen in Figure \ref{siorotate}, which shows that the slopes of the lines connecting the first two epochs are larger than those connecting the last two epochs for most of the knots. This decrease of the slopes is unlikely to be real because it is unlikely for the knots at different distances to decelerate to roughly the same value at the same time (Figure \ref{fitv}). The decrease may be due to the poorest resolution in the 2004 epoch, so that the knots in that epoch are more unresolved than the knots in other epochs. In addition, as discussed later, since SiO traces the shocks in the knots and the shocks evolve with time, the detailed structures of the knots may change, and thus the peak position of the knots may not be from the same part of the knots, as the knots propagate.

	We can now refine the velocity and the inclination angle for the jet, using the new mean proper motion measured here and the mean radial velocity of the jet measured in  \citet{2009ApJ...699.1584L}. Since the mean radial velocity was found to be $\sim$ 17 km s$^{-1}$, the velocity
of the jet is $\sim$ 115 $\pm\,50$ km s$^{-1}$ and is $\sim$ 30$\%$ lower than that measured before. The  inclination angle is $\sim$ 9$^{\circ}$, slightly larger than that ($\sim$ 6$^{\circ}$) measured before. 
	
\subsection{Knot kinematics along the jet axis}

\begin{figure}
\epsscale{0.9}
\plotone{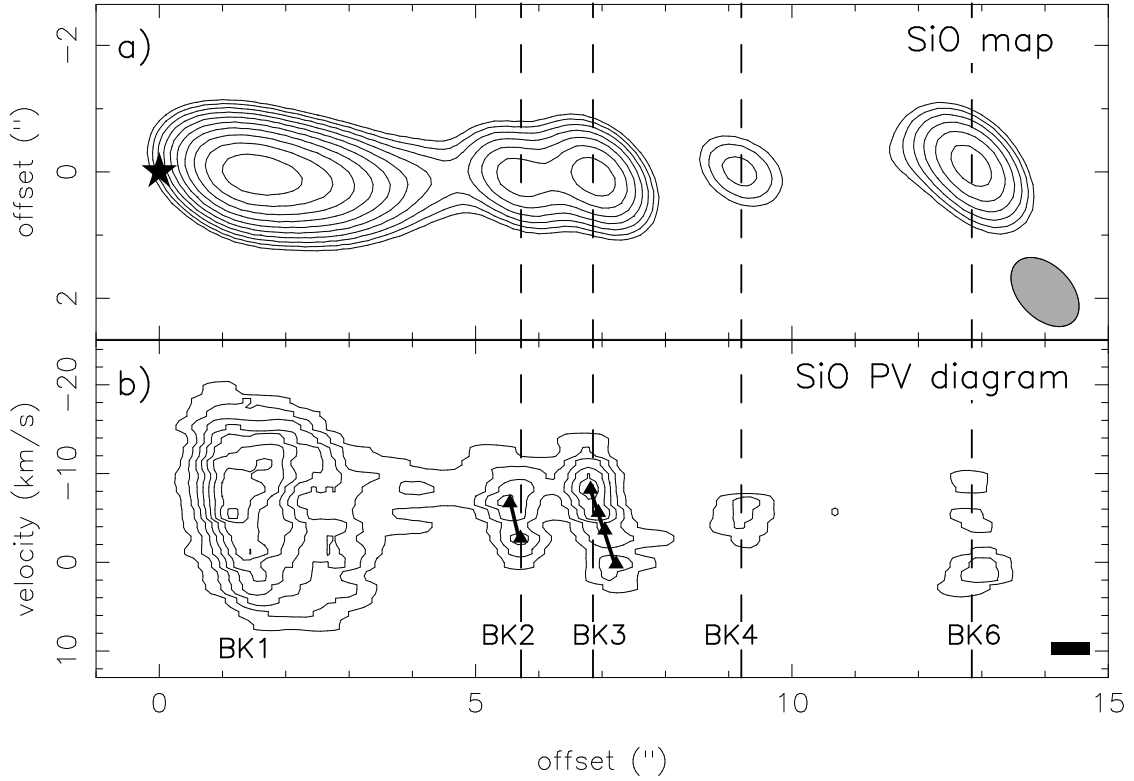}
\caption{(a) SiO map and (b) the PV diagram cut along the jet axis on the eastern side of the 2013 epoch. The system velocity is 9.2 km s$^{-1}$. The contour levels for the SiO map are the same as those in Figure \ref{siorotate}. The contour levels for the PV diagram start at 3$\sigma$ with a step of 3$\sigma$, where $\sigma\,\sim\,0.072$ Jy\,beam$^{-1}$. The asterisk symbol in SiO map indicates the position of the central source. The triangles mark the emission peaks in the PV structures. For knot BK2, the thick line connects the two peaks, and, for knot BK3, the thick line connects the peak at the highest velocity and the peak at the lowest velocity. The dashed lines indicate the peak positions of the knots in the SiO map.  The black rectangle shows the spatial and velocity resolutions of the PV diagram. \label{sioblue}}
\end{figure}

We can study the kinematics of the knots using the position-velocity (PV) diagram of the jet along the jet axis in SiO emission, as shown in Figure \ref{sioblue}. Here, we only present the PV diagram for the eastern side of the jet because the knots there have well-defined single-knot structures. We use the 2013 epoch because the PV structures of knots BK2 and BK3 are clear. These two knots are each associated with a linear PV structure consisting of two or more emission peaks (marked by the triangles in the figure), with the fast material upstream to the slow material. The solid lines connecting the peaks indicate the velocity gradients of the linear structures, and the gradient decreases from knot BK2 to BK3. Knot BK6 could also be associated with such a linear PV structure, but a higher sensitivity observation is needed to confirm it. The innermost knot, BK1, shows a much larger velocity range and a much longer structure than the other knots. It may contain several unresolved sub-knots \citep{2009ApJ...699.1584L}, showing a complicated pattern in the PV diagram.

\begin{figure}
\epsscale{.65}
\plotone{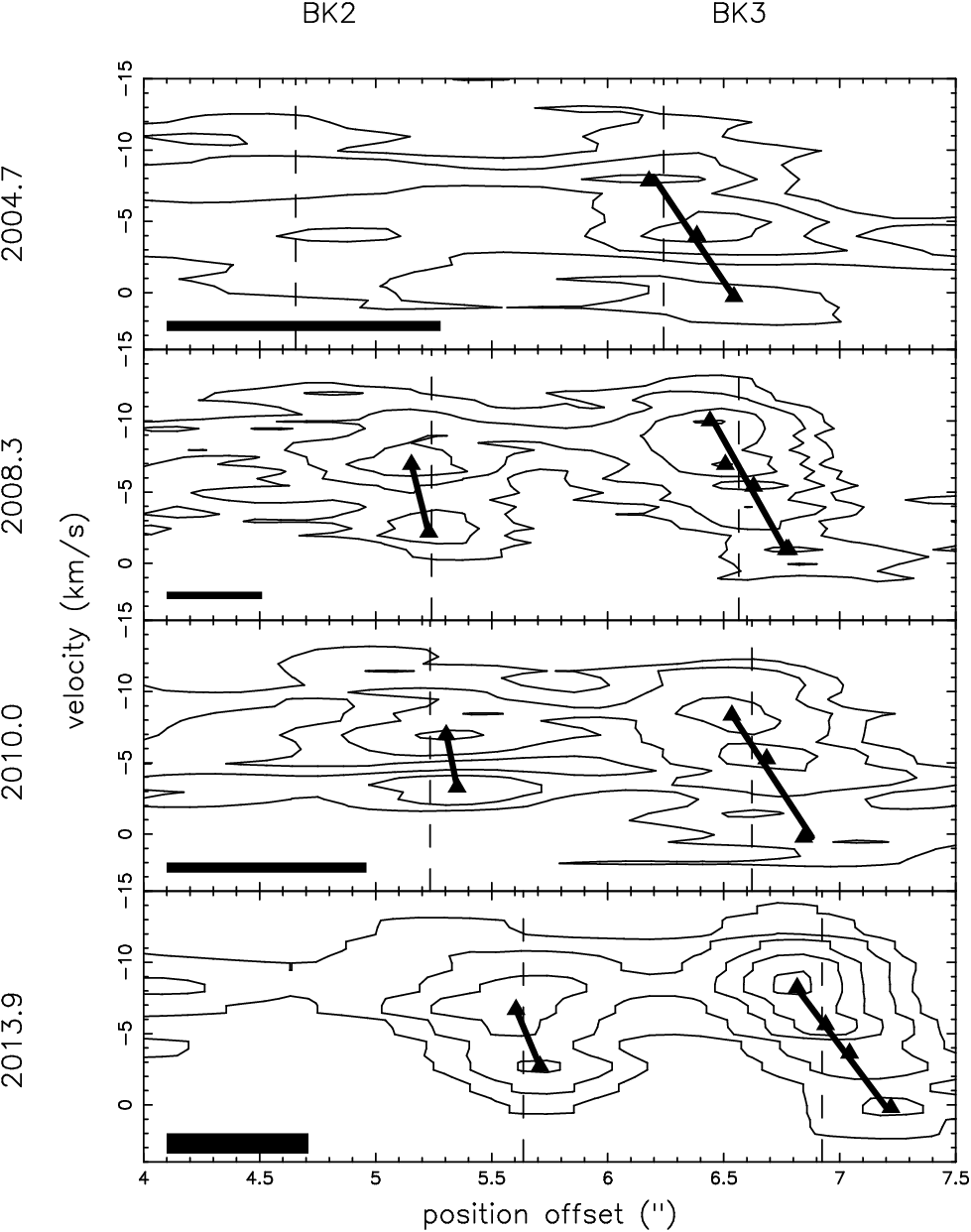}
\caption{PV diagrams cut along the jet axis for knots BK2 and BK3 in four epochs. The position offset is the distance to the central source. In the 2004 epoch, the contour levels start at 0.6 Jy beam$^{-1}$ with a step of 0.6  Jy beam$^{-1}$. In three other epochs, the contour levels start at 4$\sigma$ with a step of 3$\sigma$, and $\sigma$ is $\sim$ 0.073, $\sim$ 0.08, and $\sim$ 0.072 Jy beam$^{-1}$ for 2008, 2010, and 2013 epoch, respectively.  The triangles mark the emission peaks. As in Figure \ref{sioblue} (b), for knot BK2, the thick lines connect the two peaks, and, for knot BK3, the thick lines connect the peak at the highest velocity and the peak at the lowest velocity.\label{gradient}}
\end{figure}

For knots BK2 and BK3, the linear PV structures are also seen in earlier epochs in 2008 and 2010, as shown in Figure \ref{gradient} (presented with the original resolutions as listed in Table 1). Such a linear structure is also seen in the 2004 epoch in knot BK3, but not in knot BK2 due to poor resolution. In the figure, the solid lines show the velocity gradients of the linear structures by connecting the emission peaks. As can be seen, the gradients decrease with time from the 2004 to 2013 epochs as the knots propagate. This decrease in the gradients is likely due to the time evolution of the shocks to be discussed later. The dashed lines mark the peak positions of the knots in the SiO maps shown in Figure \ref{siorotate}, and they are located at different positions on the PV structures at different epochs. This would introduce an uncertainty in our proper motion measurement based on the emission peaks as discussed in the previous section.

\section{Discussion}
\subsection{Knot formation and internal shocks}

The formation mechanism for a chain of knots in the jet is still undetermined. In HH 211, the knots have been found to be located inside outflow cavities \citep{2007ApJ...670.1188L}.  Here, their proper motions are found to be roughly the same (even though there could be a slight difference on the opposite sides of the jet, as discussed in Section 3.2) 
and independent of the distance from the central source (Figure \ref{fitv}), indicative of no obvious deceleration of the knot motion along the jet axis. This result supports the assumption that the knots are indeed located inside the cavities and thus are not decelerated due to interaction with the ambient medium. This result also suggests that the underlying jet velocity is roughly constant and that the knots are formed due to an intrinsic property of the jet.

Two possible mechanisms have been proposed for the formation of a chain of knots in the jet.
The first mechanism is that of episodic ejections of dense jet material. \citet{2004ApJ...606..483L} have studied and ruled out this mechanism for the knots in the jets of planetary nebulae. Their simulations can also be applied to the knots of protostellar jets. In their simulations, the ejection velocity was assumed to be constant, and episodic ejections of dense jet material produce a chain of knots propagating away from the source at constant velocity, as seen in our observations. However, if this is the case, the fast material would appear downstream of the slow material in the knots \cite[see Figure 3c in][] {2004ApJ...606..483L}, inconsistent with our observations that show  the fast material upstream of the slow material. In addition, previous CO maps also show a continuous structure of the jet extending out from the source to knot BK2 on the east and to knot RK2 on the west \cite[see Figure 1c in][] {2010ApJ...713..731L}. Since CO traces roughly the column density of the material, the jet is continuous near the source due to continuous ejection. Therefore, the knots here are unlikely to be produced by episodic ejections of jet material.

\begin{figure}
\epsscale{1}
\plotone{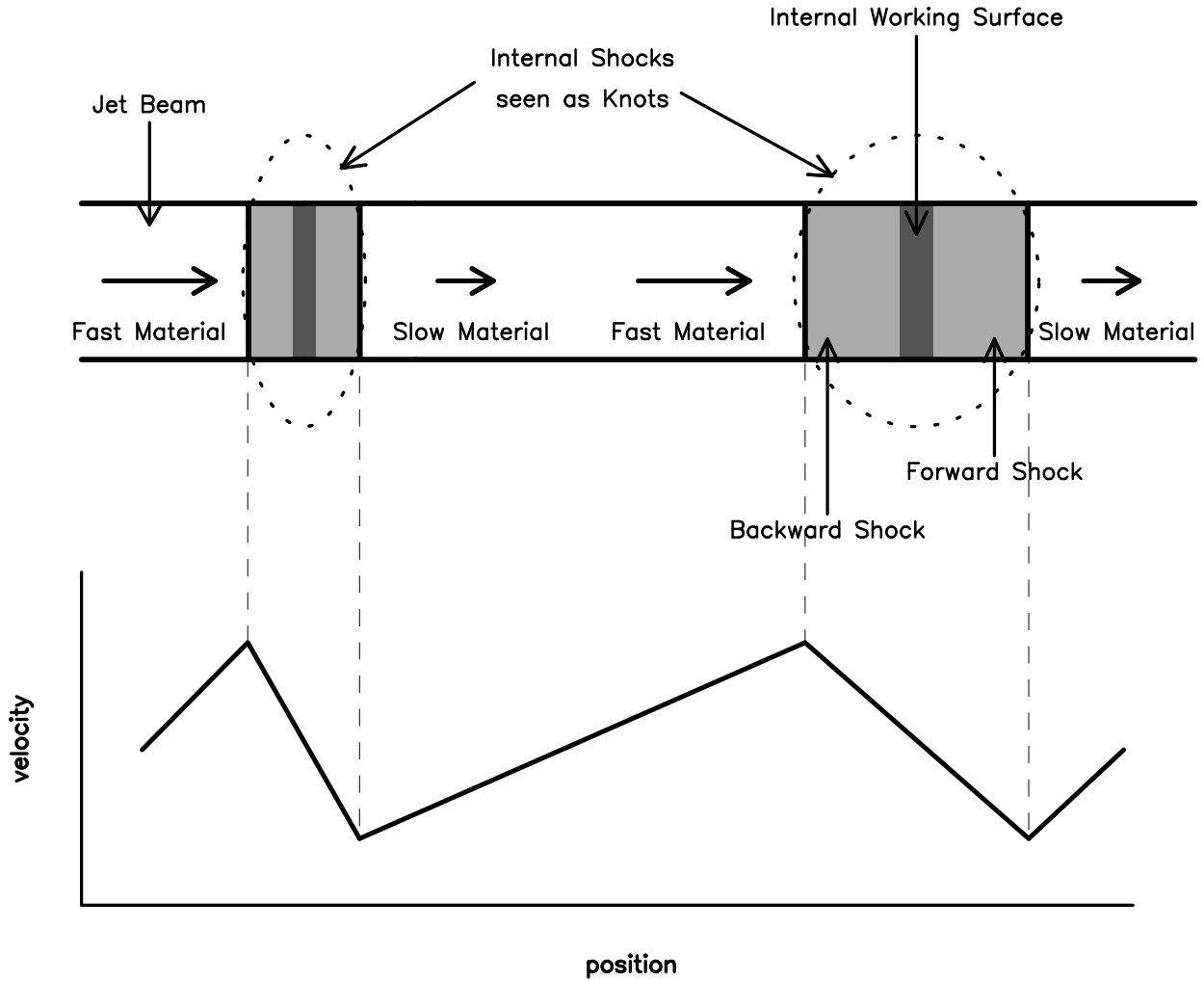}
\caption{Schematic diagram showing how fast and slow jet material interact to form internal shocks seen as knots. An internal shock consists of a backward shock, a forward shock, and an internal working surface. The bottom part of the figure shows a simplified velocity-position relationship, illustrating how the velocity and the velocity gradient in the knot change with position \citep{1990ApJ...364..601R,1993ApJ...413..210S,1997A&A...318..595S,2004ApJ...606..483L}. \label{shock}}
\end{figure}

The second possible mechanism is a small periodical variation in ejection velocity. Extensive simulations have been performed to study this mechanism by, e.g., \citet{1990ApJ...364..601R}, \citet{1993ApJ...413..210S}, \citet{1997A&A...318..595S}, and \citet{2004ApJ...606..483L}.  In those simulations, the ejection velocity of the jet is assumed to have a small periodical and sinusoidal variation, producing a chain of knots propagating away from the central source at a roughly constant velocity, as seen in our observations.  In this case, as the fast jet material catches up with the slow jet material, an internal shock region consisting of two shocks (a backward shock at high velocity and a forward shock at low velocity) and an internal working surface in between are formed, as shown in Figure \,\ref{shock}. Since SiO is known to be a good shock tracer \citep{1997A&A...321..293S}, the knots seen here in SiO are expected to trace such shock regions. In the simulations,
the velocity structure for a newly formed shock region is roughly linear, with the fast material upstream to the slow material. Therefore,
knots BK2 and BK3, which show such velocity structures, can be newly formed shock regions. The velocity gradient of the PV structure of knot BK2 in 2008 epoch is steep, further supporting knot BK2 to be a newly formed shock region.

An internal shock region does not form immediately near the source because it takes time for the fast material to catch up with the slow material.
In the simulations, the jet density is usually assumed to be constant (although some with a small variation). In this case, the jet will appear to be continuous before the internal shock region fully forms.
As mentioned earlier, previous CO maps containing both shocked and non-shocked material, already show such a continuous structure of the jet extending out from the source to knot BK2 \cite[see Figure 1c in][] {2010ApJ...713..731L}, supporting this scenario. 
Here in SiO, knot BK1 is elongated and extended to knot BK2 (Figure \ref{siorotate}), and thus may trace a region of the continuous structure where an internal shock is still forming. The fact that knot BK1 has a larger velocity range than those of knots BK2 and BK3 also supports this possibility \cite[see, e.g., the simulation results in Figure 8c of][] {2004ApJ...606..483L}. 
To further support this possibility, we estimated the distance for an internal shock to fully form, using equation 3 in \cite{1990ApJ...364..601R}. Assuming that the jet has a velocity of $v_j \sim$ 115 km s$^{-1}$ (as derived earlier), and that the velocity variation has an amplitude of $\Delta v \sim$ 30 km s$^{-1}$ (as inferred from the velocity range of knot BK1 in Figure \ref{sioblue}) and a period of $P \sim$ 20 years (since the interknot spacing is $\sim$ 2\arcsec), then the distance for an internal shock to fully form is $\sim v_j \cdot \frac{v_j}{\Delta v/(P/2)} \sim$ 1.4 $\times$ 10$^{16}$ cm ($\sim$ \arcsa{3}{3}). This distance is in between knots BK1 and BK2, supporting that knot BK1 traces a region where an internal shock is still forming and that knot BK2 traces a newly formed internal shock region.

In the simulations, as the newly formed internal shocks propagate away from the central source, they
expand, causing a decrease in the velocity gradient, as shown in Figure \ref{shock} of this paper and Figure 8c of \citet{2004ApJ...606..483L}. 
This expansion can also account for the observed decrease in the velocity gradient with time as the knots propagate, and the observed decrease in the velocity gradient from knot BK2 to BK3.
This further supports that knots BK2 and BK3 are indeed the newly formed internal shocks. 
In knot BK2, the two peaks in the PV structures of knot BK2 can be identified as two shocks (backward and forward). In knot BK3, more than two peaks are seen in the PV structures. In this case, the one at the highest velocity and the one at the lowest velocity can trace the two shocks, and the ones in between may trace the unresolved internal working surface (higher resolution observations are needed to confirm this).

In summary, although episodic ejections of jet material can produce a chain of knots as seen in the observations, they seem to have difficulty in accounting for the velocity structure of the knots and the continuous structure of the jet near the source. On the other hand, a small variation in ejection velocity of the jet with a constant jet density can account for the major features seen in our observations. Therefore, the mass-loss rate of the jet is likely to be roughly constant and the knots trace the internal shocks in the jet produced by a small velocity variation.

\subsection{The Mass-loss rate and accretion rate}
With the new jet velocity, we can refine the mass-loss rate of the jet and the accretion rate toward the central source derived before in \citet{2009ApJ...699.1584L}. The mass-loss rate of the two-sided jet is now $\dot{M}_{j} \sim 1.2\times10^{-6}$ M$_{\bigodot}$ yr$^{-1} $, about 30\% lower than that derived before. Thus, the mechanical luminosity of the jet becomes 
\[ L_{j} = \frac{1}{2}\dot{M}_{j}v_{j}^{2} \sim 1.3\,\rm{\,L_{\bigodot},} \]
using the mean jet velocity, $v_{j}$, of $\sim$ 115 km s$^{-1}$. This mechanical luminosity is about one third of the bolometric luminosity of the source \cite[$L_{bol} \sim 3.6$ L$_{\bigodot}$,][] {2005ApJS..156..169F}.

Assuming that both the bolometric luminosity of the source and the mechanical luminosity of the jet come from the accretion toward the source, we can derive the accretion rate of the source. If the central source is a single protostar with a total mass of $M_{*} \sim$ 60 M$_{\rm{Jup}}$ \citep{2005ApJS..156..169F,2009ApJ...699.1584L} and a stellar radius of $R_{*} \sim$ 2 R$_{\bigodot}$ \citep{1988ApJ...332..804S,2008ApJ...676.1088M}, the accretion rate is $\dot{M}_{a} \sim (L_{bol}+L_{j})R_{*}/GM_{*} \sim 5.3\times10^{-6}\rm{\,\,M_{\bigodot}\,\,yr^{-1}}$, about 38\% lower than that derived before. Therefore, the mass-loss rate is $\sim$ 23\% of the accretion rate, still consistent with those predicted in the jet launching models \citep{2000prpl.conf..789S,2007prpl.conf..277P}.

The central source could be a binary because of the reflection-symmetric wiggle of the jet \citep{2010ApJ...713..731L}. With the new jet velocity of $\sim$ 115 km s$^{-1}$, the mass of the jet source (primary) is found to have $M_{j} \sim 40 \,\rm{M_{\rm{Jup}}}$, and the mass of the secondary source (companion) is $M_{c} \sim 20\, \rm{M_{\rm{Jup}}}$. Since the bolometric luminosity scales with the mass of source, the jet source can have two thirds of the total bolometric luminosity. In this case, the accretion rate toward the jet source would be $\sim 5.9\times10^{-6}\rm{\,\,M_{\bigodot}\,yr^{-1}}$. Thus, the mass-loss rate is $\sim$ 20\% of the accretion rate, also consistent with those predicted in the jet launching models \citep{2000prpl.conf..789S,2007prpl.conf..277P}. Only one jet is seen from the binary system probably because the mass of the secondary source is small.

\section{Conclusions}
We have used 4 epochs (2004, 2008, 2010, and 2013) of SMA archive data to study the properties of the HH 211 jet in SiO (J=8-7). The jet consists of a chain of knots and shows a small-scale wiggle, as seen before. The conclusions are the following: 

1. The wiggle structures are similar in different epochs, and they can be fitted by an orbiting jet source model that includes a position shift due to proper motion of the jet. This confirms that the wiggle structures propagate away from the central source and thus are due to an orbital motion of the jet source, not due to an instability.

2. Six knots are used for proper motion measurements. The proper motion of the knots are roughly the same ($\sim$ 114 km s$^{-1}$) and independent of the distance from the central source, indicative of no obvious deceleration of knot motion along the jet axis. This supports the assumption that the knots are indeed located inside outflow cavities. 

3. Knots BK2 and BK3 have well-defined linear velocity structures with the fast jet material upstream to the slow jet material. The velocity gradients decrease from knot BK2 to BK3. In addition, for each knot, the gradient decreases with time, as the knots propagate away from the central source. These results are expected if these two knots trace internal shocks produced by a small periodical variation of ejection velocity of the jet. The emission peaks seen in the velocity structures may trace different parts of the internal shocks.
 
4. With the newly derived jet velocity, the mass-loss rate is $\dot{M}_{j} \sim 1.2\times10^{-6}\rm{\,\,M_{\bigodot}\,yr^{-1}}$,  which is $\sim$ 20\% of the accretion rate ($\sim 5.3\times10^{-6}\rm{\,\,M_{\bigodot}\,yr^{-1}}$), consistent with that of current jet launching models.

\acknowledgments
K.-S.J. and C.-F.L. acknowledge grants from the National Science Council of Taiwan (NSC 101-2119-M-001- 002-MY3, MOST 104-2119-M-001-015-MY3) and the Academia Sinica (Career Development Award). We also thank Anthony Moraghan for useful discussion.

\appendix
%\section{Appendix material}

\begin{table}
\begin{center}
\caption{Array configuration, resolution, and sensitivity in each epoch of the HH 211 observations.}
\begin{tabular}{cccc}
\\
\hline
\hline
epoch (yr) & array configuration & resolution & sensitivity\\
\hline
2004.7 & compact and extended & \arcsa{1}{28} $\times$ \arcsa{0}{84} & 0.47 K\\
2008.3 & extended and very extended & \arcsa{0}{46} $\times$ \arcsa{0}{36} & 0.25 K\\
2010.0 & extended & \arcsa{1}{00} $\times$ \arcsa{0}{68} & 0.27 K\\
2013.9 & extended & \arcsa{0}{70} $\times$ \arcsa{0}{51} & 0.18 K\\

\hline
\end{tabular}
\end{center}
\end{table}

\end{document}